\newcounter{myctr}
\def\myitem{\refstepcounter{myctr}\bibfont\noindent\ifnum\themyctr>9\else\phantom{0}\fi\hangindent17pt\themyctr.\enskip}

\documentclass{ws-ijqi}

\usepackage{amssymb}

\begin{document}

\markboth{E. Milotti et al.}
{Exploring quantum vacuum with low-energy photons}

\catchline{}{}{}{}{}

\title{EXPLORING QUANTUM VACUUM WITH LOW-ENERGY PHOTONS}

\author{E. MILOTTI\footnote{
Corresponding author: Edoardo Milotti,  INFN - Sez. di Trieste and Dip. di Fisica, Universit\`a di Trieste, via A. Valerio 2, I-34127 Trieste, Italy. e-mail: milotti@ts.infn.it}, F. DELLA VALLE}
\address{INFN - Sez. di Trieste and Dip. di Fisica, Universit\`a di Trieste\\ via A. Valerio 2, I-34127 Trieste, Italy}

\author{G. ZAVATTINI, G. MESSINEO}
\address{INFN - Sez. di Ferrara and Dip. di Fisica, Universit\`a di Ferrara\\ via Saragat 1, Blocco C, I-44122 Ferrara, Italy}

\author{U. GASTALDI, R. PENGO, G. RUOSO}
\address{INFN - Lab. Naz. di Legnaro\\ viale dell'Universit\`a 2, I-35020 Legnaro, Italy}

\author{D. BABUSCI, C. CURCEANU, M. ILIESCU, C. MILARDI}
\address{INFN, Laboratori Nazionali di Frascati, \\CP 13, Via E. Fermi 40,
I-00044, Frascati (Roma), Italy}

\maketitle


\begin{abstract}
Although quantum mechanics (QM) and quantum field theory (QFT) are
highly successful, the seemingly simplest state -- vacuum --
remains mysterious. While the LHC experiments are expected to clarify basic questions on the structure of QFT vacuum,  much can still be done at lower energies as well. For instance, experiments like PVLAS try to reach extremely high sensitivities,
in their attempt to observe the effects of the interaction of visible or near-visible photons with intense magnetic fields -- a process which becomes possible
in quantum electrodynamics (QED) thanks to the vacuum fluctuations of the
electronic field, and which is akin to photon-photon scattering. 
PVLAS is now close to data-taking and if
it reaches the required sensitivity, it could provide important
information on QED vacuum. PVLAS and other similar experiments face great challenges as they try to measure an extremely minute effect. However, raising the photon energy
greatly increases the photon-photon cross-section,  and gamma rays
could help extract much more information from the observed light-light scattering.
Here we discuss an experimental design to measure photon-photon
scattering close to the peak of the photon-photon cross-section, that could fit in the proposed construction of an FEL facility at the Cabibbo Lab near Frascati (Rome, Italy). 
\end{abstract}


\section{Introduction}

The quantum theories of  fields rank among the most spectacularly successful theories of physics, and yet they display some evident shortcomings that show that we still do not understand some basic elements of the physical world. Certainly, the most important failing is the uncanny mismatch between the measured energy density of vacuum and the energy density of the ground level of the fundamental fields\cite{sw,la,cpt}, which is wrong by something like 120 orders of magnitude. This is the so called ``cosmological constant problem'', and several potential solutions have been devised to get rid of it, but at the moment we cannot pinpoint any of these solutions as the final one\cite{sw}.\footnote{As S. Weinberg put it, in an influential review paper on this problem\cite{sw}: ``Physics thrives on crisis. \dots Perhaps it is for want of other crises to worry about that interest is increasingly centered on one veritable crisis: theoretical expectations for the cosmological constant exceed observational limits by some 120 orders of magnitude. ''}

It is important to devise experiments that can help disentangle this complex knot. For instance, the search for supersymmetric particles at the LHC and in astroparticle experiments could lead to momentous breakthroughs. In this paper we briefly discuss an experiment in which we study the vacuum of QED at the low-energy end, and we propose an accelerator setup that could accommodate a higher-energy version of the same experiment, with far fewer noise problems. 

\section{Light-light or light-field interaction in the low-energy limit}

In the absence of matter, Maxwell's equations can be obtained from the classical electromagnetic Lagrangian density ${\cal L}_{\rm Cl}$ (in S.I. units)
\begin{equation}
{\cal L}_{\rm Cl} = \frac{1}{2\mu_{\rm 0}}\left(\frac{E^{2}}{c^{2}}-B^{2}\right)
\end{equation}
In this case the superposition principle holds, and light-light scattering and other nonlinear electromagnetic effects in vacuum are excluded.

In 1936, after the introduction of Dirac's equation for electrons and Heisenberg's Uncertainty Principle, Euler and Heisenberg\cite{QED} derived a Lagrangian density which leads to electromagnetic nonlinear effects even in vacuum. For photon energies well below the electron mass and fields much smaller than their critical values, $B \ll B_{\rm crit} = {m_{e}^{2}c^{2}}/{e \hbar} = 4.4\cdot10^{9}$ T, $E \ll E_{\rm crit} = {m_{e}^{2}c^{3}}/{e \hbar} = 1.3\cdot10^{18}$~V/m, the Euler-Heisenberg Lagrangian correction can be written as
\begin{equation}
{\cal L}_{\rm EH} = \frac{A_{e}}{\mu_{\rm 0}}\Bigg[\Big(\frac{E^{2}}{c^{2}}-B^{2}\Big)^{2}+7\Big(\frac{\mathbf{E}}{c}\cdot\mathbf{B}\Big)^{2}\Bigg]
\end{equation}
where $\mu_{0}$ is the magnetic permeability of vacuum and
\begin{equation}
A_{e} = \frac{2}{45\mu_{0}}\frac{\alpha^{2}\mathchar'26\mkern-10mu\lambda_e^{3}}{m_{e}c^{2}} = 1.32\cdot10^{-24} \;{\text T}^{-2}
\label{Ae}
\end{equation}
with $\mathchar'26\mkern-10mu\lambda_e$ being the Compton wavelength of the electron, $\alpha={e^2}/{(4\pi\epsilon_0\;\hbar c )}$ the fine structure constant, $m_e$ the electron mass, $c$ the speed of light in vacuum. 

\begin{figure}[pb].
\centerline{\psfig{file = 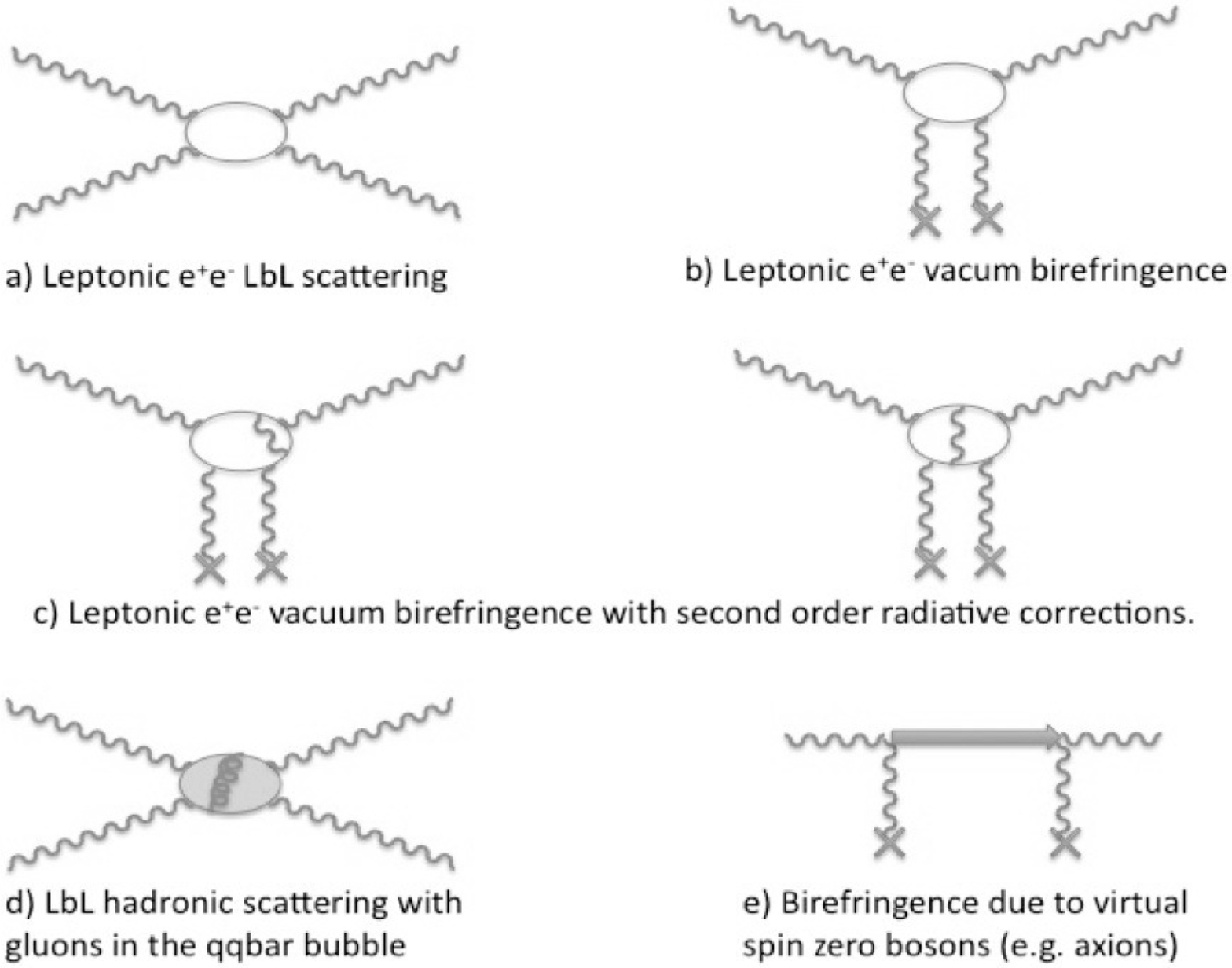, angle = -90, width = 10 cm}}
\caption{Feynman diagrams for four field interactions.}
\label{4photon}
\end{figure}

This Lagrangian correction allows four-field interactions and can be represented, to first order, by the Feynman diagrams shown in Figure \ref{4photon} a) and b).
 Figure \ref{4photon} a) represents light-by-light (LbL) scattering whereas Figure \ref{4photon} b) represents the interactions of real photons with a background electric or magnetic field, leading to vacuum  birefringence (either electric or magnetic).
 
The magnetic case is specially important, because it is easier to produce a high energy density with diffuse magnetic fields rather than with electric fields. To determine the magnetic birefringence of vacuum the electric displacement vector $\mathbf D$ and magnetic intensity vector $\mathbf H$ are derived  from the total Lagrangian density $\cal L = \cal L_{\rm Cl} + \cal L_{\rm EH}$ by using the relations\cite{Adler}
\begin{equation}
 \mathbf D = \frac{\partial {\cal L}}{\partial \mathbf E} \quad {\rm and} \quad \mathbf H = - \frac{\partial {\cal L}}{\partial \mathbf B}
 \end{equation}
 From these one obtains
 \begin{subequations}
 \begin{eqnarray}
 \mathbf D &=&\epsilon_{\rm 0}\mathbf E + \epsilon_{0}A_{e}\Big[4\Big(\frac{E^{2}}{c^{2}}-B^{2}\Big)\mathbf E + 14 \Big(\mathbf E\cdot\mathbf B\Big)\mathbf B\Big] \label{D}
\\
 \mathbf H &=& \frac{\mathbf B}{\mu_{\rm 0}} + \frac{A_{e}}{\mu_{\rm 0}}\Big[4\Big(\frac{E^{2}}{c^{2}}-B^{2}\Big)\mathbf B - 14 \Big(\frac{\mathbf E\cdot\mathbf B}{c^2}\Big)\mathbf E\Big]
 \label{H}
 \end{eqnarray} 
 \end{subequations}
 From now on one proceeds as in (nonlinear) material media: it is evident that the superposition principle no longer holds, because of the nonlinear dependence of $\mathbf D$ and $\mathbf H$ with respect to $\mathbf E$ and $\mathbf B$. 
 By assuming a linearly polarized beam of light propagating perpendicularly to an external magnetic field $\mathbf B_{\rm ext}$ there are two possible configurations: light polarization parallel to $\mathbf B_{\rm ext}$ and light polarization perpendicular to $\mathbf B_{\rm ext}$. By substituting $\mathbf E = \mathbf E_{\gamma}$ and $\mathbf B = \mathbf B_{\gamma} + \mathbf B_{\rm ext}$ in (\ref{D}) and (\ref{H}) one finds the following relations for the relative dielectric constants and magnetic permeabilities
 \begin{eqnarray}
 \left\{ \begin{array}{ll}
 \epsilon_{\parallel} &= 1 + 10 A_{e}B_{\rm ext}^2\\
 \mu_{\parallel} &= 1 + 4 A_{e}B_{\rm ext}^2\\
n_{\parallel} &= 1 + 7 A_{e}B_{\rm ext}^2
\end{array} \right.
\quad
 \left\{ \begin{array}{ll}
 \epsilon_{\perp} &= 1 - 4 A_{e}B_{\rm ext}^2\\
 \mu_{\perp} &= 1 + 12 A_{e}B_{\rm ext}^2\\
n_{\perp} &= 1 + 4 A_{e}B_{\rm ext}^2
\end{array} \right.
\label{index}
 \end{eqnarray}

From these sets of equations two important consequences are apparent: the velocity of light is no longer $c$ when an external magnetic field is present, and vacuum becomes birefringent with
 \begin{equation}
 \Delta n = n_\parallel-n_\perp = 3 A_{e} B_{\rm ext}^2
 \label{birifQED}
 \end{equation}
 
Light-by-light scattering probes into quantum vacuum, however it is by no means limited to QED. In the following sections we delineate briefly how refinements and new physics could modify the QED prediction, and open a new window on QED vacuum. 
 
\subsection{Post-Maxwellian generalization}
Quantum field theories are  challenged by string theories and their offshootsand one particular prediction is that the effective Lagrangian correction should be a Born-Infeld (BI) Lagrangian $\mathcal{L}_{BI}$ (see, e.g., references \refcite{amb} and \refcite{chru} ), defined by
\begin{equation}
\mathcal{L} = \mathcal{L}_{Cl} +\mathcal{L}_{BI} = \frac{1}{2\mu_0}\left\{ \left( \frac{E^2}{c^2} -B^2 \right) +\frac{1}{b^2}\left[ \left( \frac{E^2}{c^2} -B^2 \right)^2 +4\left( \frac{\mathbf{E}}{c}\cdot\mathbf{B} \right)^2  \right] \right\}
\end{equation}
where $b$ is the maximum magnetic field strength in the BI theory, rather than the Euler-Heisenberg Lagrangian $\mathcal{L}_{EH}$. To take into account this and other possible variants, it is interesting to generalize the non linear electrodynamic Lagrangian density correction by introducing two dimensionless free parameters depending on the model, $\eta_{\rm 1}$ and $\eta_{\rm 2}$:
\begin{equation}
{\cal L}_{\rm pM} = \frac{\xi}{2 \mu_{\rm 0}}\left[\eta_{\rm 1}\left(\frac{E^{2}}{c^{2}}-B^{2}\right)^{2}+4\eta_{\rm 2}\left(\frac{\mathbf{E}}{c}\cdot\mathbf{B}\right)^{2}\right]
\end{equation}
where $\xi = 1/B_{\rm crit}^{2} = \left(\frac{e \hbar}{m_{e}^{2} c^{2}}\right)^{2}$. With such a formulation the birefringence induced by an external magnetic field is
 \begin{equation}
 \Delta n = 2 \xi\left(\eta_{\rm 2}-\eta_{\rm 1}\right) B_{\rm ext}^2
 \end{equation}
This expression reduces to (\ref{birifQED}) if $\eta_{\rm 1} = \alpha /45 \pi$ and $\eta_{\rm 2} = 7/4 \eta_{\rm 1}$. It is thus apparent that $n_{\parallel}$ depends only on $\eta_{1}$ whereas $n_{\perp}$ depends only on $\eta_{2}$. It is also noteworthy that if $\eta_{\rm 1} = \eta_{\rm 2}$, as is the case in the Born-Infeld model, then there is no magnetically induced birefringence even though elastic scattering will be present.\cite{Haissinsky,Denisov}

 \subsection{New physics}
\label{sec:3}
The QED vacuum can also be modified by other fields interacting with photons and this means that by exploring vacuum we wade into a world of possible new physics. As mentioned above, two other important hypothetical effects could also cause $n \ne 1$ in the presence of an external magnetic (or electric) field transverse to the light propagation direction. These can be due either to neutral bosons weakly coupling to two photons called axion-like particles (ALP),\cite{Petronzio,Sikivie,Gasperini,Raffelt} or millicharged particles (MCP).\cite{Gies,Ringwald,fermion,landaulev,spin0} In this second case both fermions and spin-0 particles can be treated.
\subsubsection{ALP}
Search for axions using laboratory optical techniques was experimentally pioneered by the BFRT collaboration\cite{Cameron} and subsequently continued by the PVLAS effort.\cite{HypIn,PRD,scatter} Initially, this second experiment published the detection of a dichroism induced by the magnetic field\cite{PRL} in vacuum. Such a result, although in contrast with the CAST experiment,\cite{CAST} could have been due to axion-like particles. Subsequently the result was excluded by the same collaboration\cite{PRD,scatter} after a series of upgrades to their apparatus and almost simultaneously the axion-like interpretation was excluded by two groups\cite{RizzoALP1,RizzoALP2,gammeV} in a regeneration type measurement. However, the original publication revived interest in the optical effects which could be caused by ALP's and later MCP's.

The Lagrangian densities describing the interaction of either pseudoscalar fields $\phi_{\rm a}$ or scalar fields $\phi_{\rm s}$ with two photons can be expressed as (for convenience, written in natural Heaviside-Lorentz units)
\begin{equation}
{\cal L}_{a} = \frac{1}{M_{a}}\phi_{a} \mathbf{E}\cdot \mathbf{B} \quad{\rm and}\quad
{\cal L}_{s} =  \frac{1}{M_{s}}\phi_{s} \left(E^{2}-B^{2} \right)
\label{lagalp}
\end{equation}
where $M_{a}$ and $M_{s}$ are the inverse coupling constants. 

Therefore in the presence of an external uniform magnetic field $\mathbf{B}_{\rm ext}$ a photon with electric field $\mathbf{E}_{\rm \gamma}$ parallel to $\mathbf{B}_{\rm ext}$ will interact with the pseudoscalar field whereas for electric fields perpendicular to $\mathbf{B}_{\rm ext}$ no such interaction will exist. For the scalar case the opposite is true: an interaction will exist if $\mathbf{E}_{\rm \gamma} \perp \mathbf{B}_{\rm ext}$ and will not if $\mathbf{E}_{\rm \gamma} \parallel \mathbf{B}_{\rm ext}$. When an interaction is present, an oscillation between the photon and the pseudoscalar/scalar field will exist.

For photon energies above the mass $m_{\rm a,s}$ of such particle candidates, a real production can follow. This will cause an oscillation of those photons whose polarization allows an interaction into such particles. On the other hand, even if the photon energy is smaller than the particle mass, virtual production will follow and will cause a phase delay for those photons with an electric field direction allowing an interaction.

The attenuation $\kappa$ and phase delay $\phi$ for light with polarization allowing an interaction can be expressed, in both the scalar and pseudoscalar cases\cite{Petronzio,Sikivie,Cameron}, as follows:

\begin{equation}
\kappa =2\left(\frac{B_{\rm ext}L}{4M_{a,s}}\right)^{2}\left(\frac{\sin x}{x}\right)^{2}\quad{\rm and}\quad
\phi = \frac{\omega B_{\rm ext}^{2}L}{2M_{a,s}^{2}m_{a,s}^{2}}\left(1-\frac{\sin2x}{2x}\right)
\end{equation}
where, in vacuum, $x=\frac{Lm_{a,s}^{2}}{4\omega}$, $\omega$ is the photon energy and $L$ is the magnetic field length. The above expressions are in natural Heaviside-Lorentz units whereby 1~T $=\sqrt{\frac{\hbar^{3}c^{3}}{e^{4}\mu_{0}}}= 195$~eV$^2$ and 1~m $=\frac{e}{\hbar c}=5.06\cdot10^{6}$~eV$^{-1}$.  The phase delay $\phi$ is related to the index of refraction $n$ by
\begin{equation}
\phi=k\left(n-1\right)L
\end{equation}
Therefore in the pseudoscalar case, where $n^{a}_{\parallel}>1$ and $n^{a}_{\perp}=1$, and in the scalar case, where $n^{s}_{\perp}>1$ and $n^{s}_{\parallel}=1$, the birefringence $\Delta n = n_{\parallel} - n_{\perp}$ will be
\begin{equation}
\Delta n = n_{\parallel}^{a}-1= 1-n_{\perp}^{s}= \frac{B_{\rm ext}^{2}}{2M_{a,s}^{2}m_{a,s}^{2}}\left(1-\frac{\sin2x}{2x}\right)
\label{pseudo}
\end{equation}

In the approximation for which $x\ll1$ (small masses) this expression becomes
\begin{equation}
\Delta n = n^{a}_{\parallel}-1 = 1-n^{s}_{\perp} = \frac{B_{\rm ext}^{2}m_{a,s}^{2}L^{2}}{16M_{a,s}^{2}}
\end{equation}
whereas for $x \gg 1$
\begin{equation}
\Delta n = n^{a}_{\parallel}-1 = 1-n^{s}_{\perp} = \frac{B_{\rm ext}^{2}}{2M_{a,s}^{2}m_{a,s}^{2}}
\end{equation}
The different behavior of $n^{s}_{\perp}-1$ and $n^{a}_{\parallel}-1$ with respect to $L$ in the two cases where $x\ll1$ and $x\gg1$ is interesting and leaves, in principle, a free experimental handle for distinguishing between various scenarios.

\subsubsection{MCP}
Consider now the vacuum fluctuations of particles with charge $\pm\epsilon e$ and mass $m_{\epsilon}$ as discussed by Gies and Ringwald in references \refcite{Gies} and \refcite{Ringwald}. The photons traversing a uniform magnetic field may interact with such fluctuations resulting in both a pair production if the photon energy $\omega > 2m_{\epsilon}$ and only a phase delay if $\omega < 2m_{\epsilon}$. Furthermore, either fermions or spin-0 charged bosons could exist. Since we are discussing birefringence effects only the (real) index of refraction will be considered here.

 - Dirac fermions

Let us first consider the case in which the millicharged particles are Dirac fermions (Df). As derived by Tsai in reference \refcite{fermion} the indices of refraction of photons with polarization respectively parallel and perpendicular to the external magnetic field have two different mass regimes defined by a dimensionless parameter $\chi$ (S.I. units):
\begin{equation}
\chi\equiv
\frac{3}{2}\frac{\hbar\omega}{m_{\epsilon}c^{2}}\frac{\epsilon e B_{\rm ext}\hbar}{m_{\epsilon}^{2}c^{2}}
\label{chi}
\end{equation}
It can be shown that\cite{Gies,landaulev}
\begin{equation}
n_{\parallel,\perp}^{Df}=1+I_{\parallel,\perp}^{Df}(\chi) A_{\epsilon} B_{\rm ext}^{2}
\label{ndf}
\end{equation}
with
\begin{equation}
I_{\parallel,\perp}^{Df}(\chi)=
\left\{ \begin{array}{ll}
\left[\left(7\right)_{\parallel},\left(4\right)_{\perp}\right] & \textrm { for  } \chi \ll 1 \\
-\frac{9}{7}\frac{45}{2}\frac{\pi^{1/2}2^{1/3}\left(\Gamma\left(\frac{2}{3}\right)\right)^{2}}{\Gamma\left(\frac{1}{6}\right)}\chi^{-4/3}\left[\left(3\right)_{\parallel},\left(2\right)_{\perp}\right] & \textrm{ for   }\chi\gg 1
\end{array}\right.\nonumber
\label{idf}
 \end{equation}
and
\begin{equation}
A_{\epsilon}=\frac{2}{45\mu_{0}}\frac{\epsilon^{4}\alpha^{2} \mathchar'26\mkern-10mu\lambda_\epsilon^{3}}{m_{\epsilon}c^{2}}
\end{equation}
in analogy to equation (\ref{Ae}).
In the limit of large masses ($\chi\ll1$) this expression reduces to (\ref{index}) with the substitution of $\epsilon e$ with $e$ and $m_{\epsilon}$ with $m_{e}$ in equation (\ref{ndf}). The dependence on $B_{\rm ext}$ remains the same as for the well known QED prediction.

For small masses ($\chi\gg1$) the index of refraction now also depends on the parameter $\chi^{-4/3}$ resulting in a net dependence of $n$ with $B_{\rm ext}^{2/3}$ rather than $B_{\rm ext}^{2}$. 

In both mass regimes, a birefringence is induced:
\begin{eqnarray}
&\Delta n^{Df}&=\left[I_{\parallel}^{Df}(\chi)-I_{\perp}^{Df}(\chi)\right] A_{\epsilon} B_{\rm ext}^{2}=\\
&=&\left\{\begin{array}{ll}
3 A_{\epsilon} B_{\rm ext}^{2}& \textrm{ for  } \chi \ll 1 \\
-\frac{9}{7}\frac{45}{2}\frac{\pi^{1/2}2^{1/3}\left(\Gamma\left(\frac{2}{3}\right)\right)^{2}}{\Gamma\left(\frac{1}{6}\right)}\chi^{-4/3}A_{\epsilon} B_{\rm ext}^{2}& \textrm{ for   }\chi\gg 1
\end{array}\right.\nonumber
\label{deltandf}
\end{eqnarray}

 - Spin-0 charged bosons
 
 Very similar expressions to the Dirac fermion case can also be obtained for the spin-0 (s0) charged particle case.\cite{Gies,spin0} Again there are two mass regimes defined by the same parameter $\chi$ of expression (\ref{chi}). In this case the indices of refraction for the two polarization states with respect to the magnetic field direction are
 \begin{equation}
n_{\parallel,\perp}^{s0}=1+I_{\parallel,\perp}^{s0}(\chi) A_{\epsilon} B_{\rm ext}^{2}
\label{ns0}
 \end{equation}
 with
 \begin{equation}
I_{\parallel,\perp}^{s0}(\chi)
=\left\{ \begin{array}{ll}
\left[\left(\frac{1}{4}\right)_{\parallel},\left(\frac{7}{4}\right)_{\perp}\right] & \textrm { for  } \chi \ll 1 \\
-\frac{9}{14}\frac{45}{2}\frac{\pi^{1/2}2^{1/3}\left(\Gamma\left(\frac{2}{3}\right)\right)^{2}}{\Gamma\left(\frac{1}{6}\right)}\chi^{-4/3}\left[\left(\frac{1}{2}\right)_{\parallel},\left(\frac{3}{2}\right)_{\perp}\right] & \textrm{ for   }\chi\gg 1
\end{array}\right.\nonumber
\label{is0}
 \end{equation}
 
The vacuum magnetic birefringence is therefore
\begin{eqnarray}
&\Delta n^{s0}&=\left[I_{\parallel}^{s0}(\chi)-I_{\perp}^{s0}(\chi)\right] A_{\epsilon} B_{\rm ext}^{2}=\\
&=&\left\{\begin{array}{ll}
-\frac{6}{4} A_{\epsilon} B_{\rm ext}^{2}& \textrm{ for  } \chi \ll 1 \\
\frac{9}{14}\frac{45}{2}\frac{\pi^{1/2}2^{1/3}\left(\Gamma\left(\frac{2}{3}\right)\right)^{2}}{\Gamma\left(\frac{1}{6}\right)}\chi^{-4/3}A_{\epsilon} B_{\rm ext}^{2}& \textrm{ for   }\chi\gg 1
\end{array}\right.\nonumber
\label{deltans0}
\end{eqnarray} 
As can be seen there is a sign difference in the birefringence $\Delta n$ induced by an external magnetic field in the presence of Dirac fermions with respect to the case in which spin-0 particles exist.

 \subsection{Higher order QED corrections}
Figures \ref{4photon} c) and d) show the Feynman diagrams for the $\alpha^{3}$ contribution to the vacuum magnetic birefringence. The effective Lagrangian density for this correction has been evaluated by different authors\cite{rad,bakalovrad,dunne} and can be expressed as
\begin{equation}
{\cal L}_{\rm Rad} = \frac{A_{e}}{\mu_{0}}\left(\frac{\alpha}{\pi}\right)\frac{10}{72}\left[32\Big(\frac{E^{2}}{c^{2}}-B^{2}\Big)^{2}+263\Big(\frac{\mathbf{E}}{c}\cdot\mathbf{B}\Big)^{2}\right]
\end{equation}
This Lagrangian leads to an extra correction $\Delta n_{\rm Rad}$ of the vacuum magnetic birefringence given in equation (\ref{birifQED})
\begin{equation}
\Delta n_{\rm Rad}=\frac{25\alpha}{4\pi}3 A_{e}B_{\rm ext}^{2}=0.0145 \cdot 3 A_{e}B_{\rm ext}^{2}
\end{equation}
resulting in a 1.45 \% correction.

\section{The PVLAS experiment}

The PVLAS experiment is paradigmatic of a class of low-energy experiments to measure the properties of quantum vacuum\footnote{The PVLAS experiment is the direct successor of the experiment that was originally proposed in ref \refcite{Iacopini}. Some of the authors of this paper are also members of the PVLAS Collaboration (EM, FDV, GZ,  GM, UG, RP and GR).}. The core of the PVLAS experiment is an ellipsometer capable of measuring very small ellipticities and rotations. The extremely sensitive PVLAS ellipsometer is primarily designed to measure small birefringences due to the presence of a strong dipolar magnetic field, as predicted by QED. The dipolar magnetic field $\mathbf{B}$ is produced by two strong permanent dipole magnets. The same ellipsometer can also be used to measure small rotations due to dichroism, and attributable to the same magnetic field: although QED does not predict dichroism (photon splitting is unmeasurably small), ALP's and MCP's could contribute to it. 

In this section we give a short description of PVLAS: for a full, more detailed account, see Ref.~\refcite{pvlast}.

A birefringence $\Delta n$ induces an ellipticity $\Psi$ on a linearly polarized beam of light given by
\begin{equation}
\Psi = \pi \frac{L_{\rm eff} {\Delta n}}{\lambda}\sin2\vartheta
\end{equation}
where $L_{\rm eff}$ is the effective path length within the birefringent region with birefringence $\Delta n$ and $\lambda$ is the wavelength of the light traversing it. The induced ellipticity also depends on the angle $\vartheta$ between the light polarization and the magnetic field direction. In the QED case $\Delta n$ depends quadratically on the magnetic field $\mathbf{B}$. Therefore the magnetic field region must be as long as possible, the magnetic field as intense as possible and the wavelength small. Finally the expected ellipticity must be compared to the different noise sources present and to the maximum available integration time.

Experimentally $L_{\rm eff}$ can be made very long by using a very high finesse Fabry-Perot cavity. In fact given a birefringent region of length $L$ within a Fabry-Perot cavity of finesse $\cal F$ the effective path length is $L_{\rm eff} = \frac{2 \cal F}{\pi}L$. Today finesses ${\cal F} > 400000$ can be obtained.

High magnetic fields can be obtained with superconducting magnets but it is desirable to have a time dependent field either by ramping it, thereby changing $\Delta n$ or by rotating the field direction, thereby changing $\vartheta$. This makes superconducting magnets far less appealing than permanent magnets which, today, can reach fields above 2.5 T. Furthermore permanent magnets are relatively inexpensive to buy, have no running costs and have 100\% duty cycle allowing in principle very long integration times.

As for the wavelength we work with a Nd:YAG laser emitting radiation at $1064$ nm. Frequency doubled versions exist and could double the induced ellipticity but at the moment the highest finesses have been obtained without the frequency doubling.

Lastly it is necessary to make the magnetic field time dependent to move away from DC measurements and limit $1/f$ noise. Two detection schemes exist: homodyne detection or heterodyne detection. This second technique has been adopted in the PVLAS collaboration.

\begin{figure}[pb]
\centerline{\psfig{file = 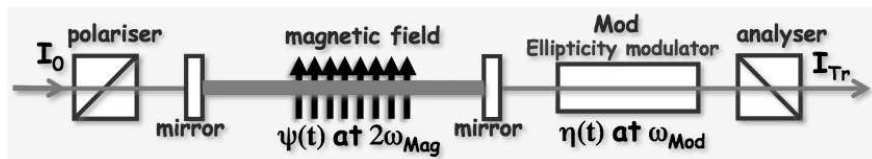, width = 12 cm}}
\caption{Scheme of the PVLAS ellipsometer.}
\label{Scheme}
\end{figure}

A scheme of the ellipsometer is shown in Figure \ref{Scheme}.
The input polarizer linearly polarizes the laser beam of intensity $I_{0}$ which then enters the sensitive region delimited by the Fabry-Perot cavity mirrors. The laser is phase locked to this cavity thus increasing the optical path length within the magnetic field by a factor $2{\cal F}/\pi$ where ${\cal F}$ is the finesse of the cavity. After the cavity the laser beam passes through a photo-elastic ellipticity modulator (PEM) which adds a known time dependent ellipticity $\eta (t)$ to the beam. This modulator ellipticity adds to the ellipticity $\Psi (t)$ acquired within the magnetic field region. After the PEM the beam passes through the analyzer which selects the polarization perpendicular to the input polarization. A photodiode detects $I_{\rm Tr}$ and its Fourier spectrum is then analyzed.

\subsection{Numbers}
A low-energy experiment like PVLAS seeks to detect a tiny effect, and to appreciate the difficulties involved in this task it is useful to present some numerical values. These numbers must be interpreted in the light of the tiny  vacuum magnetic birefringence due to the Euler-Heisenberg Lagrangian. 

In the apparatus under construction, we will have a total magnetic field length $L = 1.6$ m with a field intensity $|\mathbf B_{\rm ext} | = 2.5$ T resulting in $B^{2}_{\rm ext}L = 10.5$ T$^2$m. At present, in the test apparatus, we are running with a finesse ${\cal F} =  240000$. In the past we have reached a maximum finesse value of ${\cal F} = 414000$. Such values have also been published by other authors. We can therefore assume a value of ${\cal F} =  400000$ for our calculation. Putting these numbers together leads to
\begin{equation}
\Psi_{\rm PVLAS} = 2{\cal F}\frac{3 A_{e} L B_{\rm ext}^{2}}{\lambda} = 3\cdot 10^{-11}
\end{equation}
Assuming a maximum integration time $T_{\rm max} = 10^{6}$ s and a signal to noise ratio SNR = 1 implies that the sensitivity must be 
\begin{equation}
s_{\rm PVLAS} < \Psi_{\rm PVLAS}\sqrt{T_{\rm max}} = 3\cdot 10^{-8} \;\frac{1}{\sqrt{\rm Hz}}
\end{equation}

As discussed in Ref.~\refcite{scatter} the ultimate shot noise ellipticity sensitivity limit with the heterodyne technique depends only on the current generated in the photodiode. Assuming the power output from the cavity $I_{\rm Tr} = 5$~mW and the quantum efficiency of the diode $q = 0.7$ A/W.
\begin{equation}
s_{\rm shot} = \sqrt{\frac{e}{2I_{\rm Tr}q}} \sim 5\cdot10^{-9}\; \frac{1}{\sqrt{\rm Hz}} 
\end{equation}  

At present our sensitivity with $I_{\rm Tr} = 5$ mW is about $s_{\rm Exp} \sim 3\cdot10^{-7}\;\frac{1}{\sqrt{\rm Hz}}$ at the frequency of interest. Work is still underway to detect and suppress all the noise sources.

A careful treatment of the polarization optics in the ellipsometer shows that the power output of the final detection photodiode, including losses, is\cite{pvlast,towardsQED}
\begin{equation}
I_{\rm Tr}(t)=I_{\rm out}\Bigg|\imath\alpha(t)+\imath\eta(t)+\imath\left(\frac{1+{\it r}^{2}}{1-{\it r}^{2}}\right)\psi\sin{2\vartheta}\Bigg|^{2}
\label{Iout}
\end{equation}
This expression is at the basis of the ellipsometer in the PVLAS apparatus.\cite{dicindotto}
Small ellipticities add up algebraically and the Fabry-Perot multiplies the single pass ellipticity $\psi\sin{2\vartheta}$, generated within the cavity, by a factor $({1+{\it r}^{2}})/({1-{\it r}^{2}})\approx{2{\cal F}}/{\pi}$, where ${\cal F}$ is the finesse of the cavity. The ellipticity signal to be detected is therefore $\Psi = ({2{\cal F}}/{\pi})\psi\sin{2\vartheta}$.

In the PVLAS apparatus we modulate $\eta(t)=\eta_{0}\cos(\omega_{\rm Mod}t+\theta_{\rm Mod})$ and the magnetic field direction is rotated at an angular velocity $\Omega_{\rm Mag}$. A Fourier analysis of the power $I_{\rm Tr}(t)$ of equation (\ref{Iout}) results in four main frequency components each with a definite amplitude and phase. These are reported in table \ref{components}.
\begin{table}[!ht]
\caption{Intensity of the frequency components of the signal after the analyzer $\mathbf{A}$.}
\begin{tabular}{c|c|c|c}
\hline\noalign{\smallskip}
Frequency & Fourier component & Intensity/$I_{0}$ & Phase\\
\noalign{\smallskip}\hline\noalign{\smallskip}
$DC$ & $I_{\rm DC}$ & $\sigma^2+\alpha_{\rm DC}^2+\eta_{0}^{2}/2$ & $-$\\
$\omega_{\rm Mod}$ & $I_{\omega_{\rm Mod}}$ & $2\alpha_{\rm DC}\eta_{0}$ & $\theta_{\rm Mod}$\\
$\omega_{\rm Mod}\pm2\Omega_{\rm Mag}$ & $I_{\omega_{\rm Mod}\pm2\Omega_{\rm Mag}}$ & $\eta_{0}\frac{2{\cal F}}{\pi}\psi$ & $\theta_{\rm Mod}\pm2\theta_{\rm Mag}$\\
$2\omega_{\rm Mod}$ & $I_{2\omega_{\rm Mod}}$ & $\eta_{0}^{2}/2$ & $2\theta_{\rm Mod}$\\
\noalign{\smallskip}\hline
\end{tabular}
\label{components}
\end{table}

The presence of a component at $\omega_{\rm Mod}\pm2\Omega_{\rm Mag}$ in the signal identifies an induced ellipticity within the Fabry-Perot cavity. Furthermore the phase of this component must satisfy the value in table \ref{components}.

\subsection{Other experiments}

PVLAS is not alone in its quest, and  there are other experiments that are currently being set up to measure the magnetic birefringence of vacuum\cite{Brodin,exawatt,Luiten1,Luiten2,bmv,ni,pugnat,heinzl}. The basic technique remains the same, although the implementation can differ markedly, as is the case of the planned BMV experiment\cite{bmv}. BMV utilizes a very strong, pulsed magnetic field: the field is much higher than in PVLAS, and this leads to a correspondingly higher birefringence, however the duty cycle is much reduced, at the level of about $10^{-6}$. It is interesting to compare the two approaches, and we note that in magnetic birefringence measurements the important experimental parameter is 
\begin{equation}
\epsilon = \frac{\mathcal{F}}{\lambda \sqrt{D_t}} \int_0^{L_B} B^2 dL
\end{equation}
where $\mathcal{F}$ is the finesse of the Fabry-Perot cavity (substantially equivalent to
the number of passes through the magnetic field region), $L_B$ is the magnetic  field length, $\lambda$ is the wavelength used and $D_t$ is the duty cycle. 
The corresponding actual running values (4 hours running per 24 hour day: liquid helium filing in the morning, running in the afternoon, resting at night) for PVLAS and expected (5 pulses per hour, 24 hours per day) values for BMV are respectively $\mathcal{F}_\mathrm{PVLAS} = 37000$, $\left[ \int_0^{L_B} B^2 dL\right]_\mathrm{PVLAS} = 5.3$ T$^2$m, $\lambda = 532$ nm, $[D_t]_\mathrm{PVLAS} \approx 0.17$ and $\mathcal{F}_\mathrm{BMV} = 481000$, $\left[ \int_0^{L_B} B^2 dL\right]_\mathrm{BMV} = 25$ T$^2$m, $\lambda = 1064$ nm, $[D_t]_\mathrm{BMV} = 5.6 \cdot 10^{-6}$. These values result in $\epsilon_\mathrm{PVLAS} = 6.2\cdot 10^{10}$ T$^2$ and $\epsilon_\mathrm{BMV} = 27\cdot 10^{10}$ T$^2$ respectively. Thus, although BMV has both a higher finesse and higher value of $\int_0^{L_B} B^2 dL$,  the extremely low duty cycle completely eliminates these advantages. The slightly better optical sensitivity of BMV (the signal is at higher frequencies and is therefore less subject to $1/f$ noise) partially compensates this loss due to the small duty cycle with a final result that the total running time necessary to reach the same sensitivity as PVLAS is
approximately the same. Ultimately the success of either scheme shall depend on the noise level attained and on the ability to control all the systematic effects. 

\section{Light-light scattering cross-section and ``high-energy'' tests}

We have seen in the previous sections that a low-energy measurement is extremely difficult because of the smallness of the QED effect and because of the large background noise, that must be suppressed to a very high degree to carry out the actual measurement, as well as because of systematic effects that must all be kept in check. However the photon-photon cross section is a fast-growing function of the photon energy, and a higher-energy measurement would be an interesting alternative approach. In this section we briefly consider the basics of photon-photon scattering, and how it relates to the underlying QED structure, and put forward some suggestions for possible  experiments at a dedicated photon-photon scattering facility. 

\subsection{Photon-photon scattering cross-section}
The photon-photon scattering cross-section was first considered by Halpern\cite{Halp}, and the low-energy cross section was first calculated by Euler and Kockel\cite{EK}. The complete cross-section was  first calculated by Akhieser et al\cite{akh}. Later, it was  thoroughly studied in a series of papers by Karplus and Neuman\cite{KN1,KN2}, and by De Tollis\cite{DT1,DT2}. It turns out that a quantity that plays a very important role is the so called {\em vacuum polarization tensor} $G_{\mu\nu\lambda\sigma}^{(\kappa)}$, which depends on the photon four-momenta, and is completely symmetric with respect to indices and momenta, and is also divergenceless and  P-invariant: 
\begin{equation}
G_{\mu\nu\lambda\sigma}^{(\kappa)}(k^{(1)},k^{(2)},k^{(3)},k^{(4)}) = G_{\mu\nu\lambda\sigma}^{(\kappa)}(-k^{(1)},-k^{(2)},-k^{(3)},-k^{(4)})
\end{equation}
Here the upper index denotes photons, and conventionally 1 and 2 denote the incoming photons; $\kappa = m c^2/\hbar c $ is the reciprocal length related to the mass of the fermion (electron) in the  loop shown in Figure \ref{4photon}a. In order to obtain meaningful results, the vacuum polarization tensor must be regularized 
\begin{equation}
G_{\mu\nu\lambda\sigma} = \lim_{M\rightarrow\infty} \left[ G_{\mu\nu\lambda\sigma}^{(\kappa)} - G_{\mu\nu\lambda\sigma}^{(M)}  \right]
\end{equation}
The differential cross-section can be computed directly from the vacuum polarization tensor, and  in the CM reference frame where $\mathbf{p}=\mathbf{k}^{(1)}=-\mathbf{k}^{(2)}$, $\mathbf{q}=\mathbf{k}^{(3)}=-\mathbf{k}^{(4)}$, $\omega=|\mathbf{p}|=|\mathbf{q}|$,   it is
\begin{equation} 
\frac{d\sigma}{d\Omega} = \frac{\alpha^2 r_e^2}{4\pi^2 k^2} | M |^2
\end{equation}
where $r_e$ is the classical electron radius; here $M$ represents generically the polarization dependent amplitudes that are directly related to the vacuum polarization tensor
\begin{equation}
M_{\lambda_1 \lambda_2 \lambda_3 \lambda_4}(\Omega) = \frac{1}{4} e_\mu^{\lambda_1} e_\nu^{\lambda_2} e_\lambda^{\lambda_3} e_\sigma^{\lambda_4}G_{\mu\nu\lambda\sigma}(\mathbf{p},\omega,-\mathbf{p},\omega,\mathbf{q},\omega,-\mathbf{q},\omega)
\end{equation}
When we consider the scattering of unpolarized photons, we must replace $|M|^2$ with its average over all the incoming photon polarizations and sum over all outgoing photon polarizations. At energies less than about 0.7 MeV, the  CM differential cross-section -- averaged over initial polarizations and summed over final polarizations -- is
\begin{equation}
\label{approxcs}
\frac{d\sigma}{d\Omega} \approx \frac{139 \alpha^4}{(180 \pi)^2}\frac{(\hbar \omega)^6}{(m c^2)^8} \left( 3+\cos^2\theta \right)^2
\end{equation}

\begin{figure}[pb]
\centerline{\psfig{file = 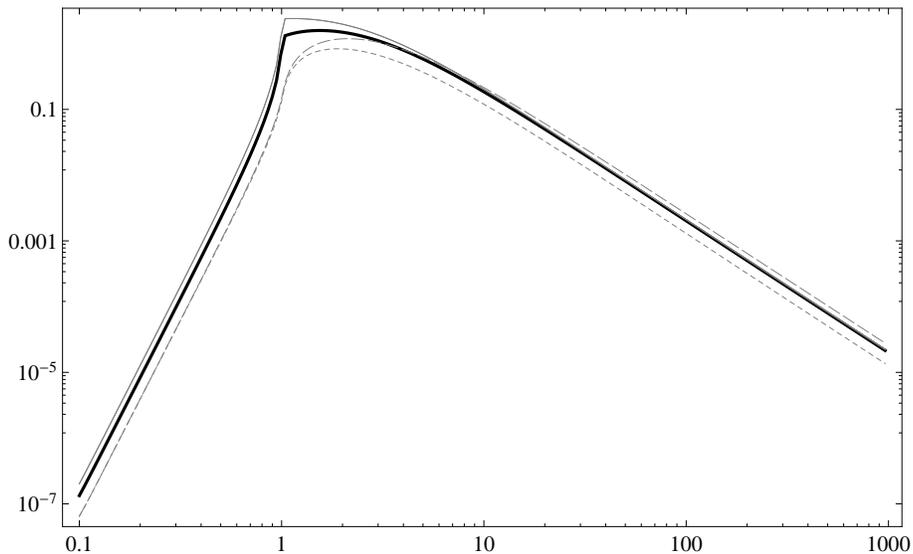, width = 12 cm}}
\caption{Total photon-photon scattering cross-section ($\mu$barn) vs. energy (MeV) . The thick black line is the unpolarized cross-section, the dotted line is the cross-section for photons that have both polarization perpendicular to the scattering plane, the dashed line is the cross--section for photons that have both polarization parallel to the scattering plane, and finally the thin solid line is the cross--section for photons that have one polarization parallel and the other perpendicular to the scattering plane.}
\label{cs}
\end{figure}

It is interesting to remark that the cross-section depends directly on the vacuum polarization tensor and that the regularization procedure plays an all-important role. Indeed, if one forgoes regularization, the cross-section turns out wrong: this was  recently evidenced by a preprint in arXiv\cite{wrong} and by a paper that rectified the wrong claims in the preprint\cite{LaC}. The relevance of regularization here involves some deep issues, as evidenced by Jackiw\cite{jack}, who distinguished three classes of radiative corrections: those that diverge and do absolutely need regularization, those that do not diverge and yield definite results without regularization, and those that do not diverge, but lead to undefined results unless they are regularized. Jackiw conjectured that
``if the form of the radiative correction is such that inserting it into the bare Lagrangian would interfere with symmetries of the model or would spoil renormalizability, then the radiative result will be finite and uniquely fixed. Alternatively, if modifying the bare Lagrangian by the radiative correction preserves renormalizability and retains the symmetries of the theory, then the radiative calculation will not produce a definite result -- it is as if the term in question is already present in the  bare Lagrangian with an undetermined strength, and the radiative correction adds a further undetermined contribution. With this criterion, the radiatively induced $g-2$ Pauli term in QED and the photon mass in the Schwinger model are unique \dots''. These considerations place the tests of $g-2$ in a different class with respect to photon-photon scattering and show that the measurement of the photon-photon cross-section is not just a test of QED as it is, but a deep test of the regularization procedure as well. 

The approximate expression (\ref{approxcs}) shows that the CM cross-section is a very fast-growing function of photon energy, and close to 1 MeV, it is about 36 orders of magnitude higher than in the visible photon range. 

It is also important to note that the ability of setting the initial photon polarization gives a partial access to the individual polarization dependent amplitudes. This is illustrated in Figure \ref{cs} which shows the cross-section for unpolarized photons, as well the cross-section for photons that are polarized parallel and perpendicular to the scattering plane. 

A photon-photon scattering experiment with photon energies in the 0.5-0.8 MeV would thus lead to an important check of our understanding of QED vacuum. The cross-section is not small, and an experiment would be feasible with current technology.

\section{A facility dedicated to photon-photon scattering}

\begin{figure}[pb]
\centerline{\psfig{file = 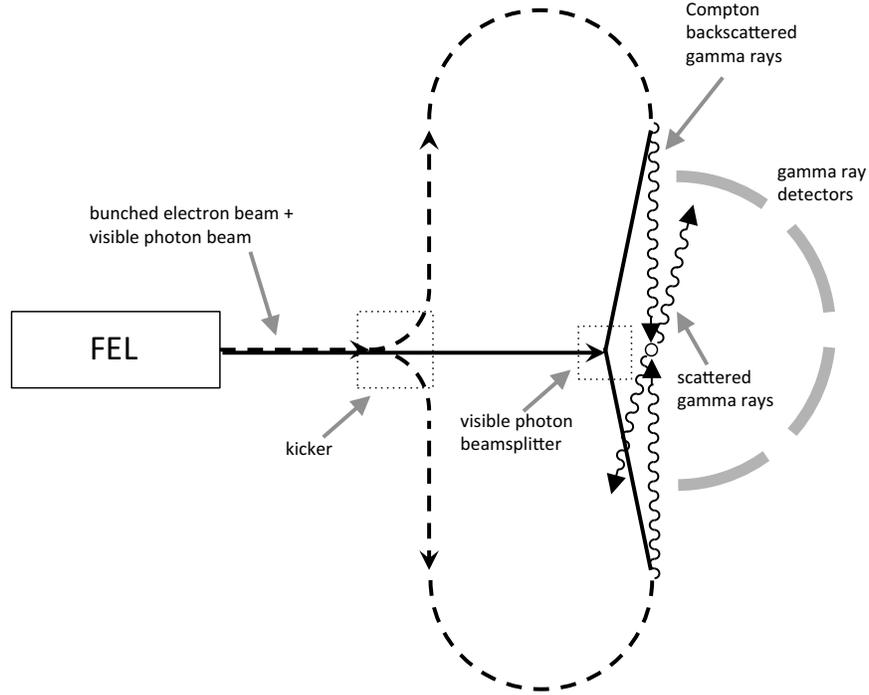, width = 12 cm}}
\caption{Scheme of a photon-photon scattering facility similar to the one proposed in references \protect\refcite{B1} and \protect\refcite{B2}. }
\label{FEL}
\end{figure}

\begin{figure}[pb]
\centerline{\psfig{file = 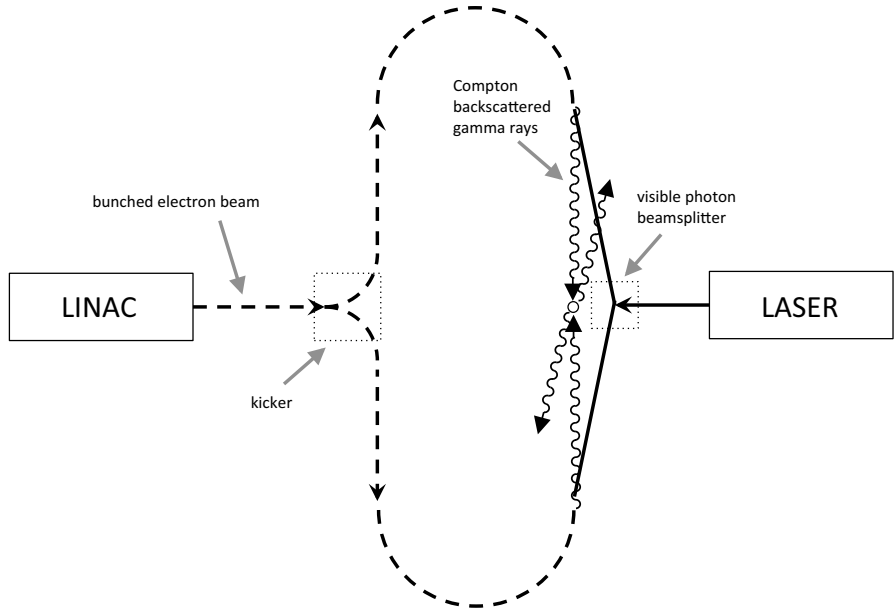, width = 12 cm}}
\caption{Scheme of a photon-photon scattering facility that utilizes an electron linac and a powerful laser, with a Compton backscattering scheme like in reference  \protect\refcite{chen}. }
\label{LINAC}
\end{figure}

A facility dedicated to photon-photon scattering has already been proposed a few years ago by Becker, McIver and Schlicker\cite{B1,B2}, and now we believe that the advances in accelerator and light-source technology make such a facility feasible and within budget of many national funding agencies. The basic idea involves an electron accelerator, such an FEL source, and an intense photon beam -- in the original proposal the very photon beam produced by the optical FEL source. By splitting both the electron beam and the photon beams to produce pairs of counter-propagating beams, and Compton backscattering to raise the energy of the photon beams, it is possible to produce a photon-photon collider. These variants are  illustrated in figures \ref{FEL} and \ref{LINAC} . 

According to references \refcite{B1} and \refcite{B2}, the total scattering rate per electron in the scheme shown in figure \ref{FEL} is
\begin{equation}
\label{CBrate}
\frac{\Gamma}{T} =  2 \alpha \nu^2 \omega_q f(\nu) = \frac{4\pi  \alpha c \nu^2}{\lambda_q}  f(\nu) 
\end{equation}
where $\omega_q$ is the angular frequency of the FEL radiation, $\lambda_q$ is the wavelength,  the $\nu$ parameter is given by the expression
\begin{equation}
\nu^2 = \frac{r_0^2}{\alpha}\frac{1}{\pi c (\hbar c)} \lambda_q^2 I
\end{equation}
$r_0$ is the classical electron radius, and $I$ is the FEL beam irradiance. Moreover, the $f(\nu)$ function is defined by the integral
\begin{equation}
f(\nu) =2\int_0^{(2\gamma)^{-1}} \left[ 1- 2\frac{(\gamma\theta)^2 (1+\nu^2)}{[1+\nu^2+(\gamma\theta)^2]^2} \right] \frac{\theta \; d\theta}{[1+\nu^2+(\gamma\theta)^2]^2} 
\end{equation}
where the upper integration limit sets the useful cone opening angle of the emitted Compton-backscattered beam. 

From equation (\ref{CBrate}) one finds that the number of backscattered photons per electron microbunch is 
\begin{equation}
N_\gamma = \left[ \left( \frac{\Gamma}{T}\right) \frac{\Delta t }{2} \right]  N_e^{\mathrm{eff}}
\end{equation}
where $\Delta t$ is the corresponding duration of the micropulse, and where $N_e^{\mathrm{eff}}$ is the number of electrons in the bunch that can actually scatter the FEL photons, and which is found by a simple scaling relation between the FEL and the electron beam cross-sectional areas
\begin{equation}
N_e^{\mathrm{eff}} = \frac{z^2}{d^2} \frac{N_e}{2}
\end{equation}
where the factor 2 now comes from splitting the electron beam in two, $N_e$ is the number of electrons in the microbunch, $z^2$ is the photon beam cross-sectional area and $d^2$ is the electron beam cross sectional area (here we assume $z < d$).
Finally the number $N$ of photon-photon scattering events per microbunch can be estimated from 
\begin{equation}
\label{Nevents}
N = \sigma \frac{N_\gamma^2}{A} 
\end{equation}
where $\sigma$ is the photon-photon scattering cross-section discussed above. 

Equations (\ref{CBrate}-\ref{Nevents}) are important in designing variants of the original scheme of Becker et al.\cite{B1,B2}. Now, the planned Cabibbo Lab infrastructure (close to the INFN Frascati Laboratories, near Rome), includes powerful LINACs and possibly a hard-X-ray FEL\cite{des1,LNF}, and it could be modified to incorporate a photon-photon scattering facility. Given the high energy of the LINACs, only photons scattered at an angle shall be usable for the photon-photon scattering experiment. In particular, recalling the relationship\cite{chen} 
\begin{equation}
E_x = \frac{(1-\beta \cos\theta_1)E_L}{1-\beta\cos\theta+[1-\cos(\theta-\theta_1)]E_L/mc^2\gamma} 
\end{equation}
between the incoming photon energy $E_L$ and the outgoing photon energy $E_x$ in the Compton backscattering process (in the original electron rest frame), where $\theta_1$ is the angle between the incoming electron and photon beams, and $\theta$ is the angle between the incoming electron beam and the scattered photon, we find that $\theta_1$ should be close to 0.12 radians to match the peak of the photon-photon scattering cross-section.
The angle between the electron and the photon beams helps in designing the experiment; what else do we gain in a high-energy facility? Unfortunately there is not much to gain from the higher energy: in the initial electron rest frame the Klein-Nishina cross-section is (see, e.g., reference \refcite{chen})
\begin{equation}
\frac{d\sigma}{d\Omega} = \frac{r_0^2}{2} R^2 \left( R+ \frac{1}{R} + \cos^2\theta_{ER}  \right)
\end{equation}
where $R$ is the ratio between outgoing and incoming photon energy in the rest frame, and $\theta_{ER} $ is the scattering angle in the rest frame. Since $R \le 1$ in the rest frame, we see that the cross section has a very weak dependence on photon energy, and not much can be expected moving to a higher photon energy. However the Cabibbo Lab complex is designed to achieve a high intensity as well: this is just where we can gain, and in the next months we plan to complete a design study with these optimization constraints in mind.

\section{Conclusions}

We believe that the physics case for a photon-photon physics both at low and at higher energies is very strong. The detection of the magnetic birefringence of vacuum or the observation of photon-photon scattering would be a powerful test of QED and of the regularization procedure, and it could give us important insights on the nature of quantum vacuum.  We are convinced that this is a strong motivation for  the construction of {\it ad hoc} beam lines in an accelerator complex like the Cabibbo Lab.


\begin{thebibliography}{0}

\bibitem{sw} S. Weinberg, Rev. Mod. Phys. {\bf 61} (1989) 1.
\bibitem{la} L. Abbott, Sci. Am. {\bf 258} (1988) nr.5, 106.
\bibitem{cpt} S. M. Carroll, W. H. Press, and E. L. Turner, Annu. Rev. Astron. Astrophys. {\bf 30} (1992) 499.
\bibitem{QED}
H. Euler and B. Kochel, Naturwiss. {\bf 23} (1935) 246.\\
W. Heisenberg and H. Euler, Z. Phys. {\bf98} (1936) 718.\\
V. S. Weisskopf, Kgl. Danske Vid. Sels., Math.-fys. Medd. {\bf14} (1936) 6.\\
J. Schwinger, Phys. Rev. {\bf82} (1951) 664.
\bibitem{Adler}
R. Baier and P. Breitenlohner, Acta Phys. Austriaca {\bf25} (1967) 212.\\
R. Baier and P. Breitenlohner, Nuovo Cimento {\bf47} (1967) 261.\\
S. L. Adler, Ann. Phys. {\bf67} (1971) 559.\\
Z. Bialynicka-Birula and I. Bialynicki-Birula, Phys. Rev. D {\bf2} (1970), 2341.
\bibitem{amb} J. Ambjorn, Y. M. Makeenko, G. W. Semenoff, R. J. Szabo, JHEP 0302 (2003) 026.
\bibitem{chru} D. Chrus\'ci\'nski, Phys. Rev. D {\bf 62} (2000) 105007.
\bibitem{Haissinsky}
J. Ha\"issinski {\it et al.} Phys. Scr. {\bf74} (2006) 678-681.
\bibitem{Denisov}
V. I. Denisov {\it et al.}, Phys. Rev. D {\bf69} (2004) 066008.
\bibitem{Petronzio}
L. Maiani  {\it et al.}, Phys. Lett. B{\bf 175} (1986) 359.
\bibitem{Sikivie}
P. Sikivie, Phys. Rev. Lett. {\bf 51} (1983) 1415.
\bibitem{Gasperini}
M. Gasperini, Phys. Rev. Lett. {\bf 59} (1987) 396. 
\bibitem{Raffelt}
G. Raffelt and L. Stodolsky, Phys. Rev. D {\bf37} (1998) 1237.
\bibitem{Gies}
M. Ahlers {\it et al.}, Phys. Rev. D {\bf75} (2007) 035011.
\bibitem{Ringwald}
H. Gies {\it et al.}, Phys. Rev. Lett. {\bf97} (2006) 140402.
\bibitem{fermion}
W. y. Tsai and T. Erber, Phys. Rev. D {\bf12} (1975) 1132.
\bibitem{landaulev}
J. K. Daugherty {\it et al.}, Astrophys. J. {\bf273} (1983) 761.
\bibitem{spin0}
C. Schubert, Nucl. Phys. {\bf B585} (2000) 407.
\bibitem{Cameron}
R. Cameron {\it et al.}, Phys. Rev. D {\bf47} (1993) 3707.
\bibitem{HypIn}
D. Bakalov {\it et al.}, Hyperfine Interactions {\bf114} (1998) 103.\\
D. Bakalov {\it et al.}, Quantum Semicl. Opt. {\bf 10} (1998) 239.
\bibitem{PRD}
E. Zavattini {\it et al.}, Phys. Rev. D {\bf77} (2008) 032006.
\bibitem{scatter}
M. Bregant {\it et al.}, Phys Rev. D {\bf 78} (2008) 032006.
\bibitem{PRL}
E. Zavattini {\it et al.}, Phys. Rev. Lett. {\bf96} (2006) 110406.
\bibitem{CAST}
K. Zioutas {\it et al.}, Phys. Rev. Lett. {\bf94} (2005) 121301.
\bibitem{RizzoALP1}
C. Robilliard {\it et al.}, Phys. Rev. Lett. {\bf99} (2007) 190403.
\bibitem{RizzoALP2}
M. Fouch\'e {\it et al.}, Phys. Rev. D {\bf78} (2008) 032013.
\bibitem{gammeV}
A. S. Chou {\it et al.}, arXiv:0710.3783\\
A. S. Chou {\it et al.}, Phys. Rev. Lett. {\bf 100} (2008) 080402.
\bibitem{rad}
V. I. Ritus, Zh. Eksp. Teor. Fiz {\bf69} (1975) 1517 [Sov. Phys. JETP {\bf42}
(1975) 774].
\bibitem{bakalovrad}
D. Bakalov, INFN/AE-94/27 (1994) SIS publication LNF
\bibitem{dunne}
G. Dunne, arxiv:hep-th/0406216v1
\bibitem{Iacopini}
E. Iacopini and E. Zavattini, Phys. Lett. B {\bf85} (1979) 151.
\bibitem{pvlast} 
G. Zavattini {\it et al.}, Int., J. Mod. Phys. A {\bf 27} (2012) 1260017.
\bibitem{towardsQED}
F. Della Valle {\it et al.}, Optics Comm. {\bf 293} (2010) 4194.
\bibitem{dicindotto}
G. Zavattini {\it et al.}, Applied Physics B {\bf83} (2006) 571.

\bibitem{Brodin}
E. Lundstr\"om {\it et al.}, Phys. Rev. Lett. {\bf 96} (2006) 083602.
\bibitem{exawatt}
D. Tommasini {\it et al.}, Phys. Rev. A {\bf 77} (2008) 042101.
\bibitem{Luiten1}
A. N. Luiten and J. C. Petersen, Phys. Lett. A {\bf330} (2004) 429.
\bibitem{Luiten2}
A. N. Luiten and J. C. Petersen, Phys. Rev. A {\bf70} (2004) 033801.
\bibitem{bmv}
R. Battesti {\it et al.}, Eur. Phys. J. D {\bf46}, (2008) 323.
\bibitem{ni}
W.-T. Ni, Chinese J. Phys. {\bf34} (1996) 962.
\bibitem{pugnat}
P. Pugnat {\it et al.}, Czech. J. Phys. A {\bf56} (2006) C193.
\bibitem{heinzl}
T. Heinzl {\it et al.}, Optics Comm. {\bf267} (2006) 318.












\bibitem{Halp}
O. Halpern, Phys. Rev. {\bf 44} (1933) 855.
\bibitem{EK}
H. Euler and B. Kockel, Naturwissenschaften {\bf 23} (1935) 246.
\bibitem{akh} 
A. Akhieser, L. Landau, and I. Pomeranchouk, Nature {\bf 138}  (1936) 206. \\
A. Akhieser, Phys. Zeit. Sowjetunion {\bf 11}  (1937). 263
\bibitem{KN1}
R. Karplus and M. Neuman, Phys. Rev. {\bf 80} (1950) 380.
\bibitem{KN2}
R. Karplus and M. Neuman, Phys. Rev. {\bf 83} (1951) 776.
\bibitem{DT1}
B. De Tollis, Nuovo Cimento {\bf 32} (1964) 757.
\bibitem{DT2}
B. De Tollis, Nuovo Cimento {\bf 35} (1965) 1182.

\bibitem{wrong}
N. Kanda, Light-Light Scattering, arXiv:1106.0592, 2011. \\
T. Fujita and N. Kanda, A Proposal to Measure Photon-Photon Scattering, arXiv:1106.0465, 2011.
\bibitem{LaC} 
Y. Liang and A. Czarnecki, Can. J. Phys. {\bf 90} (2012) 11. 
\bibitem{jack}
R. Jackiw, Int. J. Mod. Phys. B {\bf 14} (2000) 2011.


\bibitem{B1} 
W. Becker, J. K. McIver, and R. R. Schlicker, Phys. Rev. A {\bf 38} (1988) 4891.
\bibitem{B2} 
W. Becker, J. K. McIver, and R. R. Schlicker, J. Opt. Soc. Am. B {\bf 6} (1989) 1083.
\bibitem{des1} 
S. Guiducci et al., in {\it Proceedings of the IPAC2012} (2012) New Orleans, Louisiana,  20--25 May 2012 (published on {\tt http://www.jacow.org})
\bibitem{LNF} 
D. Alesini et al.: ``A possible hard X-ray FEL with the SuperB 6 GeV electron LINAC'', INFN-12-14/LNF (2012).
\bibitem{chen} 
J. G. Chen et al., Nucl. Instr. Meth. A {\bf 580} (2007) 1184.


\end{thebibliography}
\end{document}